\documentclass[11pt,reqno,a4paper]{amsart}
\usepackage[utf8]{inputenc}
\usepackage[T1]{fontenc}
\usepackage{amsmath}
\usepackage{amsthm} 
\usepackage{amssymb,mathrsfs} 
\usepackage{amsfonts}
\usepackage[normalem]{ulem}
\usepackage{graphicx}
\usepackage{dsfont}
\usepackage{color}
\usepackage[numbers,sort&compress]{natbib}
\usepackage{hyperref}
\usepackage{doi}
\usepackage[left=2.25cm,right=2.25cm, top = 3cm]{geometry}

\usepackage{amsopn}

\newcommand\eps{\varepsilon}
\newcommand\R{\mathds{R}}

\newcommand\E{\mathbb{E}}
\newcommand\LL{\mathcal{L}_\eps}
\newcommand\e{e}
\usepackage{hyperref}
\usepackage{doi}
\usepackage[numbers,sort&compress]{natbib}
\usepackage{tgtermes}
\usepackage{mathtools}
\usepackage{color}

\newtheorem{remark}{Remark}

\author{Grégoire Ferré and Tobias Grafke}

\title[Approximate optimal controls via instanton
  expansion]{Approximate optimal controls via instanton expansion for
  low temperature free energy computation}

\begin{document}

\date{\today}

\maketitle

\begin{abstract}
  The computation of free energies is a common issue in statistical
  physics. A natural technique to compute such high dimensional
  integrals is to resort to Monte Carlo simulations. However these
  techniques generally suffer from a high variance in the low
  temperature regime, because the expectation is often dominated by
  high values corresponding to rare system trajectories.  A standard
  way to reduce the variance of the estimator is to modify the drift
  of the dynamics with a control enhancing the probability of rare
  event, leading to so-called importance sampling estimators. In
  theory, the optimal control leads to a zero-variance estimator; it
  is however defined implicitly and computing it is of the same
  difficulty as the original problem. We propose here a general
  strategy to build approximate optimal controls in the small
  temperature limit for diffusion processes, with the first goal to
  reduce the variance of free energy Monte Carlo estimators. Our
  construction builds upon low noise asymptotics by expanding the
  optimal control around the instanton, which is the path describing
  most likely fluctuations at low temperature. This technique not only
  helps reducing variance, but it is also interesting as a theoretical
  tool since it differs from usual small temperature expansions (WKB
  ansatz).  As a complementary consequence of our expansion, we
  provide a perturbative formula for computing the free energy in the
  small temperature regime, which refines the now standard
  Freidlin--Wentzell asymptotics. We compute this expansion explicitly
  for lower orders, and explain how our strategy can be extended to an
  arbitrary order of accuracy. We support our findings with
  illustrative numerical examples.
\end{abstract}

\section{Introduction}

This work is concerned with the computation of free energy-like
quantities arising in statistical physics, for diffusion processes
in the low temperature and finite time
regime~\cite{e-ren-vanden-eijnden:2002, dellago2003,
weinan2005finite, vanden-eijnden-weare:2012}.  Although such
quantities are defined by integrals, the typical high dimensionality
of the problem makes numerical integration impossible, so that one
generally resorts to Monte Carlo simulation for numerical estimations.
However, naive Monte Carlo methods often fail to provide accurate
results because of the high variance of standard estimators. This
situation typically arises because the observable of interest is
dominated by large values along rare trajectories~\cite{bucklew:2004}.

There are in general two ways for reducing the variance of naive Monte
Carlo estimators when computing free energies. One is to introduce a
bias in the dynamics, so that rare trajectories become more likely
under the new dynamics~\cite{e-ren-vanden-eijnden:2002,
pham2009continuous, hartmann2012efficient, lelievre2016partial} -- a
strategy sometimes referred to as tilting. We know at a theoretical
level that there exists a control, called optimal, which provides a
zero-variance estimator.  However, for high-dimensional systems, it is
hopeless to compute this optimal control to a high degree of accuracy,
and poor approximations may deteriorate the quality of the
estimator. It is therefore an important and challenging problem to
estimate as accurately as possible, and at a reasonable computational
cost, the zero-variance control.

Another strategy is to resort to population
dynamics~\cite{grassberger2002, del2002stability, del2004feynman,
dean2009, lecomte2007numerical, brehier2018new}, another instance of
importance sampling. The idea here is to run a series of systems in
parallel, and to select the ones that realize the rare event
dominating the expectation defining the free energy. There are various
possibilities to design a selection mechanism, some provably behaving
better than others~\cite{rousset2006control, angeli2018rare,
doucet2009tutorial}. However, it is a known fact that, in
high-dimension and at low temperature, the number of replicas needed
for performing accurate computations becomes very
large~\cite{nemoto2016population, nemoto2017finite}.  Of course, it is
also possible to combine the two approaches,
see~\cite{nemoto2016population} for an example of application to long
time large deviations computations.

We focus here on the construction of approximate optimal controls
for diffusion processes in the low temperature regime.
Since different equivalent expressions
are available for the optimal control (for instance through
stochastics, partial differential equations or variational
representations), many approximation techniques have been developed,
including cross-entropy methods~\cite{zhang2014applications},
milestoning~\cite{hartmann2012efficient}, Isaacs equation~\cite{dupuis2007subsolutions},
martingale based
techniques~\cite{lie2015martingale}, model
reduction~\cite{hartmann2016model} and forward-backward stochastic
differential equations~\cite{kebiri2018adaptive}, or more recently
machine learning based algorithms~\cite{han2016deep, weinan2017deep,
han2018solving, nusken2020solving}. We anticipate already here that,
when an approximation of the optimal control is available, it is still
not obvious that the resulting estimator should actually decrease the
variance (see~\cite{sadowsky1990large, glasserman-wang:1997, asmussen2007stochastic,
guyader2020efficient} and references therein for more insight on this
subtle issue).

The first goal of this paper is to provide a simple way to construct
approximate controls that are well-suited for variance reduction of
free energy Monte Carlo computations in the small temperature
regime. We rely for this on low temperature reaction paths
(instantons)~\cite{e-ren-vanden-eijnden:2002,
  e-ren-vanden-eijnden:2004, grafke-grauer-schaefer-etal:2014,
  grafke-vanden-eijnden:2019} by building a time-inhomogeneous Taylor
expansion around such reaction paths. This is quite different from the
standard WKB (or Freidlin--Wentzell--Graham) small temperature
expansion~\cite{graham1984weak, freidlin1998random,
  bouchet2016perturbative, bouchet2016generalisation, kifer:1977}
where series run in the small temperature parameter and are defined
through solutions to partial differential equations. From a more
mathematical viewpoint, we propose an expansion of a finite noise
Hamilton--Jacobi--Bellman equation around the solution to the
characteristic equation of the associated noiseless partial
differential equation (see~\cite[Chapter~3.2]{evans2010partial}),
which is not a standard procedure to the best of our knowledge.  With
our technique, we manage to build offline (\textit{i.e.}~involving
only computations that can be done once before starting the sampling)
an approximate control that behaves well at low temperature.  This
should be put in contrast with techniques that build a precise
estimate of the control by requiring costly on the fly
updates~\cite{vanden-eijnden-weare:2012} or solving a partial
differential equation -- see for instance the interesting Isaacs
subsolution approach~\cite{dupuis2007subsolutions}.

A second output of our work is a perturbative formula
for the free energy at low temperature. Using the optimal control expansion, we 
compute correction terms to the Freidlin--Wentzell zero-order
asymptotics to the free energy, which can be estimated without
resorting to Monte Carlo simulation. In general, we believe the
expansion we propose is an interesting object to understand more
precisely from a mathematical standpoint, in particular in view of the
theory of viscosity solutions for Hamilton--Jacobi
equations~\cite{crandall1983viscosity, evans2010partial,
fleming2006controlled, pham2009continuous}.

The paper is organized as follows. Section~\ref{sec:setting} presents
our problem (Section~\ref{sec:framework}) and recalls some well-known
facts about zero variance estimators (Section~\ref{sec:IBP}) and low
temperature reaction paths (Section~\ref{sec:introinstanton}).  We
next turn to the main results of the paper, by first presenting our
approximation of the optimal control in Section~\ref{sec:gexpansion}
and then the resulting perturbative formula for the free energy in
Section~\ref{sec:Zexpansion}.  We conclude in
Section~\ref{sec:lowtemp:applications} with some numerical
applications illustrating our results. We finish with a short
discussion, pointing out limitations of our technique, and indicating
directions to addressing them.

%%%%%%%%%%%%%%%%%%%%%%%%%%%%%%%%%%%%%%%%%%%%%%%%%%%%%%%%%%%%%%%%%%%%%%%%%%%%%%%%%%%%%%%%
\section{Optimal control and low temperature limit}
\label{sec:setting}

This section presents the overall setting of the work, and recalls
some well-known facts about optimal control and low temperature
asymptotics of exponential expectations, which we use in
Section~\ref{sec:instanton} for our approximation procedure.

%%%%%%%%%%%%%%%%%%%%%%%%%%%%%%%%%%%%%%%%%%%%%%%%%%%%%%%%%%%%%%%%%%%%%%%%%%%%%%%%%%%%%%%%
\subsection{Free energy computation}
\label{sec:framework}
We consider the computation of integrals of exponential quantities for
which numerical integration is impossible and Monte Carlo estimators
typically have a large variance. Concretely, for fixed time $T>0$ and
initial condition $x_0\in\R^d$, we consider
\begin{equation}
  \label{eq:Aeps}
  A_{\eps}=\E_{x_0} \left[ \e^{\frac{1}{\eps} f(X_T^{\eps})}\right].
\end{equation}
Here, $f: \R^d \to\R$ is a smooth function and $(X_t^{\eps})_{t\geq
  0}$ is solution to the following stochastic differential equation
in~$\R^d$ (with~$d$ a positive integer standing for the physical dimension)
\begin{equation}
  \label{eq:X}
  d X_t^{\eps} = b(X_t^{\eps})\,dt + \sqrt{\eps}\sigma
  \,dB_t,
\end{equation}
where~$(B_t)_{t\geq 0}$ is an $m$-dimensional Brownian motion, the
function $b: \R^d \to \R^d$ is smooth, and $\sigma\in\R^{d\times m}$
is such that the diffusion matrix $D=\sigma\sigma^\mathrm{T} \in
\R^{d\times d}$ is positive definite ($\sigma^\mathrm{T}$ stands for
the transpose of the matrix~$\sigma$).  In~\eqref{eq:Aeps}, we denote
by~$\E_{x_0}$ the expectation with respect to all trajectories
solving~\eqref{eq:X} and starting at the initial point $x_0\in\R^d$ at
time $t=0$. Note that we could also consider a time-dependent
function~$b$ as well as a time-position dependent diffusion
matrix~$\sigma$ without additional difficulty, but restrict to this
setting for notational simplicity. The generator of the dynamics~\eqref{eq:X} reads
\begin{equation}
  \label{eq:gen}
\LL = b\cdot\nabla + \eps \frac{\sigma \sigma^\mathrm{T}}{2}:\nabla^2,
\end{equation}
where~$\,\cdot\,$ is the scalar product in~$\R^d$. The notation~$\nabla^2$
stands for the $\R^{d\times d}$-valued Hessian operator, while for two matrices
$A,B\in\R^{d\times d}$ we write $A:B = \mathrm{Tr}(A^\mathrm{T} B)$.
The differential operators~$\nabla$, $\nabla^2$ and~$\LL$ can be defined
on smooth compactly supported functions, and 
we assume in what follows that the parameters of the model allow to
define~\eqref{eq:Aeps} as a finite quantity for all $\eps >0$, see in
particular~\cite{freidlin1998random} for technical considerations.

A motivation for studying~\eqref{eq:Aeps} is the computation of the free energy
\begin{equation}
  \label{eq:Zeps}
  Z_\eps = \eps \log A_\eps
\end{equation}
in the small temperature regime\footnote{The term free energy is
  often associated with long time problems through the
  quantity
  \[
  \lim_{T\to+\infty}\, \frac1T \log \E\left[
    \e^{\int_0^T f(X_s^\varepsilon)\,ds}
    \right].
    \]
    Here, we use the terminology associated with small temperature
    problems, like in~\cite{hartmann2012efficient}.
    
  Note that we could also consider
  expectations involving a random stopping time~$\tau$, such as
\[
 \varepsilon \log \E_{x_0} \left[ \e^{\frac{1}{\eps} f(X_\tau^{\eps})}\right],
\]
or finite time-integrated quantities like
\[
 \varepsilon \log \E_{x_0} \left[ \e^{\frac{1}{\eps} \int_0^Tf(X_s^{\eps})\, ds}\right],
\]
for a finite integration time~$T>0$, when $\varepsilon \to 0$.  These
cases can be treated by appropriately modifying the computations
performed in Appendix~\ref{sec:proofhigher}. In this paper, we present
our method by considering~\eqref{eq:Aeps} and leave the modifications
needed in other cases to the interested reader. Combining long time
and small temperature asymptotics is on the other hand is a difficult
problem, see for instance~\cite{nickelsen2018anomalous} for
interesting insights.  }.  It is known by large deviations arguments
that, under mild assumptions, it
holds~\cite{freidlin1998random,dembo2010large}
\begin{equation}
  \label{eq:Zepslim}
Z_\eps\xrightarrow[\eps\to 0]{} Z^0
\end{equation}
for some finite value~$Z^0$, see Section~\ref{sec:introinstanton}
below.  The numerical computation of~$Z^0$ is one motivation for
estimating~\eqref{eq:Aeps} when $\eps\ll 1$. In a large deviations
perspective, it is also useful to compute~$Z^0$ for numerically
estimating the rate function associated to the path measure
of~$(X_t^\eps)_{t\in[0,T]}$, which is related to~$Z^0$ through
a Legendre--Fenchel transform.
We refer to~\cite{tailleur2008simulation,ferre2018adaptive} for numerical
examples in the related infinite time context.

In the regime of small temperature, the expectation in~\eqref{eq:Aeps}
is often dominated by very large values realized over rare trajectories, which
leads to large variance Monte Carlo estimators. However, we know that
the dynamics~\eqref{eq:X} can be controlled to be turned into a
zero-variance estimator of~\eqref{eq:Aeps}, as we recall now.

%%%%%%%%%%%%%%%%%%%%%%%%%%%%%%%%%%%%%%%%%%%%%%%%%%%%%%%%%%%%%%%%%%%%%%%%%%%%%%%%%%%%%%%%
\subsection{Optimal tilting on path space}
\label{sec:IBP}
We now present the modification of~\eqref{eq:X} leading to a
zero-variance estimator of~\eqref{eq:Aeps}.  These computations are
standard provided technical conditions are met, see for
instance~\cite{fleming2006controlled,pham2009continuous}.  In this
procedure, we consider the \emph{tilted process}~$(\widetilde
X_t^{\eps})_{t\geq 0}$ solution to
\begin{equation}
  \label{eq:Xtilde}
  d \widetilde X_t^{\eps} = b(\widetilde X_t^{\eps})\,dt
  +D \nabla g(t,\widetilde X_t^{\eps})\,dt
  + \sqrt{\eps}\sigma \,d B_t,  
\end{equation}
where $g:\R_+ \times \R^d\to\R$ is an arbitrary smooth function, and we call
$\nabla g$ the \emph{control}. We restrict ourselves to gradient
controls since, as shown below, the optimal control is indeed
gradient.

First we introduce the Girsanov weight $\alpha:[0,T]\times\R^d\to\R$
associated with~$g$, namely
\[
\forall\, t \geq 0,
\quad \forall\, x \in\R^d,
\quad \alpha(t,x) =\partial_t g(t,x) + \LL g(t,x) + \frac{1}{2} |\sigma\nabla g|^2(t,x).
\]
Next we define the function $\psi_\eps:[0,T]\times\R^d\to \R_+$ as
  \begin{equation}
    \label{eq:psiA}
  \psi_{\eps}(t,x) = \E_{t,x}\left[ \e^{\frac{1}{\eps} f(X_T^{\eps})} \right],
  \end{equation}
  and
  \begin{equation}
    \label{eq:gopt}
  g_{\eps}(t,x) = \eps \log\, \psi_{\eps}(t,x).
  \end{equation}
In~\eqref{eq:psiA},~$\E_{t,x}$ refers to the expectation with respect to all realizations
of the dynamics~\eqref{eq:X} started at time~$t$ from position~$x$.
Under technical conditions,~$g_\eps$ is well-defined as a solution
(at least in a weak sense~\cite{evans2010partial})
  to the following Hamilton--Jacobi--Bellman (HJB) equation (see
  Appendix~\ref{sec:proofscontrol}):
  \begin{equation}
    \label{eq:HJB}
\left\{
    \begin{aligned}
      \partial_t g_{\eps}  + \LL g_{\eps}
      + \frac{1}{2} \left| \sigma\nabla g_{\eps} \right|^2  &= 0 & \\
      g_{\eps}(T,x) & = f(x) , & \forall\,x\in\R^d.
    \end{aligned}
    \right.
  \end{equation}
We assume in what follows that~$g_\eps$ actually exists as a unique
smooth solution of~\eqref{eq:HJB} with the probabilistic
representation~\eqref{eq:psiA}-\eqref{eq:gopt}, and refer to
Section~\ref{sec:gexpansion} for more details on this
 assumption.

Then, by setting $g=g_\eps$ in~\eqref{eq:Xtilde}, the estimator
  \begin{equation}
    \label{eq:estimbias}
  A_{\eps} =\e^{g(0, x_0)} \E_{x_0} \left[ \exp\left(\frac{1}{\eps}
    \big[f(\widetilde X_T^{\eps}) - g_\eps(T,\widetilde X_T^{\eps}) \big]
    + \frac{1}{\eps} \int_0^T \alpha(t,\widetilde X_t^{\eps})\,dt
    \right)\right]
  \end{equation}
  has zero variance. Namely
  \begin{equation}
    \label{eq:zerovarianceA}
  A_{\eps}  = \psi_{\eps}(0, x).
  \end{equation}

This result is a consequence of the Feynman--Kac formula and the
Girsanov theorem, see Appendix~\ref{sec:proofscontrol} for a more
detailed argument.  A consequence of~\eqref{eq:estimbias} is
that~\eqref{eq:Aeps} can be estimated with a zero-variance
(\textit{i.e.} deterministic) estimator provided~\eqref{eq:gopt} is
known.

In general, the Monte Carlo estimator built on~\eqref{eq:estimbias} by
drawing independent trajectories distributed according
to~\eqref{eq:Xtilde} cannot be used as such for numerical
applications, because estimating~\eqref{eq:gopt} and its gradient for
all $t\geq 0$ and $x\in\R^d$ is still more difficult than solving the
initial problem of estimating~\eqref{eq:Aeps}. However, this result
serves as a guide to design approximate controls that are easier to
compute while still reducing the variance of Monte Carlo estimators
of~\eqref{eq:Aeps}. We will present in Section~\ref{sec:instanton} an
original strategy to build such approximate controls behaving well in
the small~$\eps$ regime from the low temperature asymptotics provided
by transition path theory, which is the main contribution of this
work. For this, we first need to recall the definition of the
transition path in our context, which is the purpose of the next
section.

%%%%%%%%%%%%%%%%%%%%%%%%%%%%%%%%%%%%%%%%%%%%%%%%%%%%%
\subsection{Low temperature regime and reaction path}
\label{sec:introinstanton}
Even though the control~$g_{\eps}$ is difficult to estimate in
practice, we can nevertheless have access to an \emph{instanton}, or reaction
or transition path, which stands for the zero temperature most likely
path of fluctuation for the dynamics. In the small noise limit, we
know by the Freidlin--Wentzell
theory~\cite[Section~3]{freidlin1998random} that the trajectories
of~$(X_t^{\eps})_{t\in[0,T]}$ dominating the
expectation~\eqref{eq:Aeps} concentrate exponentially fast on this
path for the uniform norm under relatively mild conditions on the
parameters of the problem.  We only recall the most important features
of the theory here, and refer to~\cite{grafke-vanden-eijnden:2019} and
references therein for more details.

The instanton is a path $(\phi_t)_{t\in[0,T]}$ taking values in~$\R^d$, assumed here to be
smooth and uniquely defined (we shall discuss more this assumption in
Remark~\ref{rem:multiple} below). In order to
provide an equation for this path, we also consider a conjugate
variable $(\theta_t)_{t\in[0,T]}$, which can be thought of as a momentum.
The reaction path $(\phi_t,\theta_t)_{t\in[0,T]}$ is then described by
the following forward-backward system of equations:
\begin{equation}
  \label{eq:instanton}
  \left\{
    \begin{aligned}{}
      &\dot \phi_t = b(\phi_t) + D \theta_t, \qquad& &\phi_0 = x_0,\\
      &\dot \theta_t = -(\nabla b)^\mathrm{T}(\phi_t) \theta_t ,& &\theta_T = \nabla f(\phi_T).
    \end{aligned}
  \right.
\end{equation}
Note that the initial condition~$x_0$ of~$\phi$
is the same as the one appearing in the definition~\eqref{eq:Aeps} of the free energy.
We insist on the fact that the instanton is defined by a deterministic system of equations, and
that the reaction path~$(\phi_t)_{t\in[0,T]}$ corresponds to a typical path whose final value
of~$f$ dominates the expectation in~\eqref{eq:Aeps}. Moreover, we mention that the
set of equations~\eqref{eq:instanton} is simply the characteristic system describing
the noiseless limit of the HJB equation~\eqref{eq:HJB}, see~\cite[Chapter~3.2]{evans2010partial}.

Finally, the set of equations defining the reaction path provides a representation of the low
temperature limit~\eqref{eq:Zepslim} of the free energy through
\begin{equation}
  \label{eq:Z0}
Z^0 =\lim_{\eps\to 0}\, Z_{\eps}=  f(\phi_T)- \tfrac12
   \int_0^T \theta_t\cdot D \theta_t\, dt.
\end{equation}
Thus, $Z^0$ can be interpreted in an optimal control sense as the
maximal value of~$f$ that can be reached under a quadratic
penalization of the momentum~\cite{hartmann2012efficient}. Not surprisingly, this
kind of asymptotics is obtained via the Girsanov theorem through
computations similar to that of Section~\ref{sec:IBP}.

We now have all the tools to present the main contributions of the paper, which are: (i) an 
approximation of the optimal control~$g_\eps$ around the reaction path, and (ii)
a resulting expansion of~$Z_\eps$ for small values
of~$\eps$.

%%%%%%%%%%%%%%%%%%%%%%%%%%%%%%%%%%%%%%%%%%%%%%%%%%%%%%%%%%%%%%%%%%%%%%%%%%%%%%%%%%%%%%%%
\section{Low temperature approximation of the optimal bias}
\label{sec:instanton}

We now present our main results. First, we build an approximation
 of the optimal control around the instanton 
in Section~\ref{sec:gexpansion}. We next deduce in Section~\ref{sec:Zexpansion} a perturbative
formula for the free energy~$Z_\eps$.

%%%%%%%%%%%%%%%%%%%%%%%%%%%%%%%%%%%%%%%%%%%%%%%%%%%%%%%%%%%%%%%%%%%%%%%%%%%%%%%%%%%%%%%%

\subsection{Expansion around the instanton}
\label{sec:gexpansion}
In order to present our expansion, we first recall that the zero-variance control
is the solution to the HJB equation~\eqref{eq:HJB} which reads in full form
\begin{equation}
  \label{eq:HJBeps}
\left\{
    \begin{aligned}
  \partial_t g_\eps  + b\cdot\nabla g_\eps +\eps
\frac{D}{2}:\nabla^2 g_\eps
+ \frac{1}{2} \left| \sigma\nabla g_\eps \right|^2  &= 0 & \\
      g_{\eps}(T,x) & = f(x) , & \forall\,x\in\R^d.
    \end{aligned}
    \right.
\end{equation}
In the zero-temperature limit $\eps\to 0$, the partial differential
equation above becomes
\begin{equation}
  \label{eq:HJBzero}
  \left\{
    \begin{aligned}
  \partial_t g^0  + b\cdot\nabla g^0 
  + \frac{1}{2} \left| \sigma\nabla g^0 \right|^2 & = 0& \\
      g^0(T,x) & = f(x) , & \forall\,x\in\R^d.
    \end{aligned}
    \right.
\end{equation}
We assume in what follows that~\eqref{eq:HJBeps} possesses a unique smooth
solution on $[0,T)\times \R^d$, which is typically the case under reasonable
assumptions by parabolic
regularity (see for instance~\cite[Section~4, Theorem~4.1]{fleming2006controlled}).
Moreover, we also assume that~\eqref{eq:HJBzero} has a unique smooth
solution. This is a more stringent assumption for which it is difficult
to provide general conditions of application. However, we know by the method
of characteristics that this assertion is valid when the final time~$T$ is small
enough~\cite[Section~3.2, Theorem~2]{evans2010partial}.
We place ourselves in this setting in this paper, and refer to Remark~\ref{rem:multiple} below
for further comments on these assumptions. Note
also that it typically holds in a weak sense that $\lim_{\varepsilon \to 0} g_{\eps} = g^0$.

Solving the characteristics system for~\eqref{eq:HJBzero} actually
relies~\cite{evans2010partial,grafke-vanden-eijnden:2019} on plugging the ansatz
\begin{equation}
  \label{eq:g1}
g^0(t,x) = \theta_t\cdot ( x - \phi_t)
\end{equation}
into~\eqref{eq:HJBzero}, which allows to derive the couple of
equations~\eqref{eq:instanton} defining the instanton.  The
definition~\eqref{eq:g1} is motivated by Lagrangian considerations in
statistical physics~\cite{grafke-vanden-eijnden:2019}, but is simply an
application of the method of characteristics for first order non-linear
partial differential equations~\cite[Chapter~3.2]{evans2010partial}. 

The main idea of this paper is to consider~\eqref{eq:g1} as the first term
of a polynomial Taylor expansion around the instanton~$(\phi_t)_{t\in[0,T]}$. This
suggests going to next order by looking for a solution of~\eqref{eq:HJBeps} in the form
\begin{equation}
  \label{eq:g2}
  g^1(t,x) = \theta_t\cdot (x - \phi_t) +\frac{1}{2} \big(x - \phi_t)
  \cdot K_t (x - \phi_t),  
\end{equation}
where $(K_t)_{t\in[0,T]}$ is a $\R^{d\times d}$-valued process to be
determined.
In what follows, we call the ansatz~\eqref{eq:g1} the
zeroth order approximation and~\eqref{eq:g2} the first order one,
because the resulting controls~$\nabla g^0$ and~$\nabla g^1$ are
respectively of zeroth and first order in~$x - \phi_t$ (see
Remark~\ref{rem:higher} below for expansions to arbitrary order). Although this
is not an expansion in powers of~$\varepsilon$, the temperature
appears implicitly through the relation
$\widetilde X_t^\varepsilon - \phi_t = \mathrm{O}(\sqrt{\varepsilon})$,
which holds when the drift is chosen accordingly. Indeed, taking for example
$g=g^0$ in~\eqref{eq:Xtilde} and assuming that
$\widetilde X_t^\varepsilon = \Phi_t + \mathrm{O}(\sqrt{\varepsilon})$ for some
path~$(\Phi_t)_{t\in[0,T]}$, then~\eqref{eq:Xtilde} becomes:
\[
\dot \Phi_t +\mathrm{O}(\sqrt{\varepsilon}) = b( \Phi_t)
+ D \nabla g^0(t,\Phi_t) + \mathrm{O}(\sqrt{\varepsilon}).
\]
Since $\nabla g^0(t,x) = \theta_t$, we observe that indeed $\Phi_t = \phi_t$ is
the instanton. As a result, one should think of $x-\phi_t$ as a quantity
of order~$\sqrt{\varepsilon}$ along a trajectory~$(\widetilde X_t^\varepsilon)_{t\in[0,T]}$
when the drift~$g$ is built from~$g^0$.

%%%
We now derive heuristically the equation satisfied by~$(K_t)_{t\in[0,T]}$ for~$g^1$
to be an approximation of~$g_\varepsilon$. For this, we note
that~\eqref{eq:HJBeps} rewrites componentwise
$
\partial_t g_\eps + b_k\partial_k g_\eps +  D_{jk} \partial_j g_\eps\partial_k g_\eps/2
 + \eps D_{jk}\partial_{jk} g_\eps /2= 0$,
where we use Einstein's notation for summation over repeated indices.
Taking the derivative twice with respect to indices
$i\in\{1,\hdots,d\}$ and $l\in\{1,\hdots,d\}$ shows that
\[
\partial_t\partial_{il}^2 g_\eps + \partial_{il}^2 b_k \partial_k g_\eps
+ \partial_i b_k \partial_{kl}^2 g_\eps
+ \partial_l b_k \partial_{ik}^2 g_\eps
+ b_k  \partial_{ikl}^3 g_\eps
+   D_{jk} \partial_{jl}^2 g_\eps\partial_{ik}^2 g_\eps
+   D_{jk}  \partial_j g_\eps\partial_{ikl}^3 g_\eps \eps \frac{D_{jk}}{2}\partial_{ijkl} g_\eps= 0.
\]
This can be written in vectorial form as
\[
  \partial_t \nabla^2 g_\eps + \nabla^2 b\cdot \nabla g_\eps +
  (\nabla b)^\mathrm{T} \nabla^2 g_\eps + \nabla^2 g_\eps
  \nabla b + (\nabla^2 g_\eps)^\mathrm{T} D \nabla^2 g_\eps + b\nabla^3 g_\eps
  + (\nabla^3 g_\eps)^\mathrm{T}
  D \nabla g_\eps +  \eps \frac{D}{2}:\nabla^4 g_\eps = 0,
\]
where the equation is evaluated at any~$(t,x)$.

Since we look for an evolution
equation for $K_t = \nabla^2 g^1 (t,\phi_t )$, we compute
\begin{equation}
  \label{eq:calculusg}
  \begin{aligned}
  \frac{d}{dt}\nabla^2 g_\eps(t,\phi_t) = & \ \partial_t \nabla^2 g_\eps
  + \dot{\phi}_t\cdot\nabla^3 g_\eps
\\ = & - (\nabla^2 b)^\mathrm{T} \nabla g_\eps - (\nabla b)^\mathrm{T} \nabla^2 g_\eps
- \nabla^2 g_\eps
\nabla b - (\nabla^2 g_\eps)^\mathrm{T} D \nabla^2 g_\eps 
\\ & - b\nabla^3 g_\eps - (\nabla^3 g_\eps)^\mathrm{T} D \nabla g_\eps
+ \dot{\phi}_t\cdot \nabla^3 g_\eps-  \eps \frac{D}{2}:\nabla^4 g_\eps,
  \end{aligned}
\end{equation}
where the right hand side is evaluated at $(t,\phi_t)$. Considering the ansatz~$g^1$ given
by~\eqref{eq:g2} to replace~$g_\eps$, we have
$\nabla g^1 (t,\phi_t ) = \theta_t$, $\nabla^2 g^1(t,\phi_t) = K_t$,
 $\nabla^3 g^1(t,\phi_t) = 0$ and  $\nabla^4 g^1(t,\phi_t) = 0$, which
 can be plugged into~\eqref{eq:calculusg} to obtain the equation
 satisfied by~$(K_t)_{t\in[0,T]}$. The final condition on~$K_T$
 can be derived similarly by differentiating twice the terminal condition in~\eqref{eq:HJBeps}.
As a consequence, in order for~\eqref{eq:g2} to
approximate~\eqref{eq:HJBeps}, $(K_t)_{t\in[0,T]}$ should be solution to
\begin{equation}
  \label{eq:EDOK}
  \left\{\begin{aligned}
&  \dot{K}_t + (\nabla b)^{\mathrm{T}} K_t + K_t^{\mathrm{T}}
  \nabla b +  \nabla^2 b \cdot \theta_t + K_t^{\mathrm{T}} D K_t   = 0,
  \\& K_T   = \nabla^2 f(\phi_T),
  \end{aligned}
  \right.
\end{equation}
where~$b$ and its derivatives are evaluated at~$\phi_t$.
A precise derivation of~\eqref{eq:EDOK} can be found in
Appendix~\ref{sec:proofhigher} via the Girsanov theorem (see in particular~\eqref{eq:hg}
and~\eqref{eq:Kproof}). Note that a solution to~\eqref{eq:EDOK} is symmetric.

In order to formalize that~$g^1$ is indeed an approximation of~$g_\eps$,
the solution to~\eqref{eq:HJBeps}, we thus consider
  the first order approximation~\eqref{eq:g2}
  where $(\phi_t,\theta_t)_{t\in[0,T]}$ is defined in~\eqref{eq:instanton}
  and~$(K_t)_{t\in[0,T]}$ satisfies~\eqref{eq:EDOK}. We also define the following
  function of time:
  \begin{equation}
    \forall\ t\in [0,T],\quad Z_\eps^1(t) = f(\phi_T) - \int_t^T \theta_s\cdot D\theta_s\,ds +
    \frac{\eps}{2}\int_t^T D:K_s\,ds.
  \end{equation}
  Then, we show in Appendix~\ref{sec:proofhigher} that,
  in the small~$\eps$ limit, for any $t\in [0,T]$ and $x\in\R^d$, it holds
  \begin{equation}
    \label{eq:gexpand}
    g_{\eps}(t,x) = g^1(t,x) +  Z^1_\eps(t) + \mathrm{o}(\eps) + \eps\mathrm{o}(x - \phi_t).
  \end{equation}
  This formula shows that~$g^1$ approximates the optimal control~$g_\varepsilon$ at
  small temperatures and around the instanton. A key ingredient of the proof
  is that, as noted above, for all time $t\geq0$,
  it holds $\widetilde X_t^{\eps} - \phi_t = \mathrm{O}(\sqrt \varepsilon)$
  (see~\eqref{eq:Xtildehigh} in Appendix~\ref{sec:proofhigher}), so the
  approximation is valid for the process~\eqref{eq:Xtilde} tilted by~$g^1$.

  Equation~\eqref{eq:EDOK} is an instance of  algebraic Riccati
equation~\cite{lancaster1995algebraic}, which
is an interesting feature compared to the more standard instanton presented in
Section~\ref{sec:introinstanton}.
Riccati equations recurrently appear in optimal control
theory~\cite{jurdjevic1997geometric}, so it is no surprise to encounter such an equation in
our approximation procedure (here a linearization) of the optimal control. The original
feature, we believe, is the fact that all the objects in the approximation are centered around
the zero-temperature instanton. Indeed, we
insist on the fact that our construction~\eqref{eq:g2} \emph{is not a WKB ansatz}
since it is not an expansion in the temperature parameter~$\eps$ (see Remark~\ref{rem:higher}
below). Actually, since
the tilted process controlled by~$g^1$
satisfies~$\widetilde X_t^\eps - \phi_t=\mathrm{O}(\sqrt{\eps})$,
this dependency in the temperature is hidden in the expansion around the instanton.

From a numerical perspective, we will use~$g^1$ as an ansatz
for the optimal control~$g_\eps$ defined in~\eqref{eq:gopt}.
Note that most techniques
relying on optimal control strategies strive to estimate~$g_{\eps}(t,x)$
and its gradient for all time~$t$ and position~$x$, which is very difficult and computationally
costly in practice~\cite{vanden-eijnden-weare:2012}.
Here, we can construct \emph{offline} an approximation of this optimal control, which
is a polynomial expansion whose coefficients depend on time only, which drastically reduces
the computational cost of the procedure.
However, since the construction relies on a small temperature expansion, we
expect this approximation to reduce the variance only in a low temperature regime -- a fact
confirmed by the numerical simulations below.

\begin{remark}[Higher order expansion and relation to WKB ansatz]
\label{rem:higher}
It is of course possible to push our method to an approximation of
order $M>1$ for~$g_\eps$ through
\begin{equation}
  \label{eq:gM}
g^M(t,x) = \sum_{k=1}^{M+1} T_k(t) \odot \big(x - \phi_t \big)^{\otimes k},
\end{equation}
where~$\otimes k$ stands for the~$k^{\mathrm{th}}$ order tensorization
of a $d$-dimensional vector,~$\odot$ the~$k^{\mathrm{th}}$ order
contraction and, for each $k\geq 1$, $T_k$ is a time
dependent~$k^{\mathrm{th}}$ order tensor. Comparatively, an expansion
in the temperature parameter~$\varepsilon$, sometimes called
Wentzell--Kramers--Brillouin (WKB) expansion, would read, up to
order~$M$,
\[
g(t,x) = \sum_{k=0}^{M} (\sqrt{\varepsilon})^k u_k(t,x),
\]
where each~$u_k$ is solution to a PDE~\cite{fleming1992asymptotic}.
Since $\widetilde X_t^\eps - \phi_t
=\mathrm{O}(\sqrt{\eps})$, the expansion~\eqref{eq:gM} looks like a WKB
expansion in powers of~$\sqrt{\eps}$ around the tilted process although this
parameter does not appear explicitly. Moreover, we do not need to solve any PDE
since we work with ordinary differential equations at the process level.
This allows faster numerical computations and the simple derivation of a
perturbative formula for the free energy~\eqref{eq:Zeps}, as presented in
Section~\ref{sec:Zexpansion} below. In some sense, the expansion we propose
can be thought of as a Taylor, polynomial expansion version of the standard
WKB series.

With the notation~\eqref{eq:gM}, we have seen above that~$T_1(t) = \theta_t$ and~$T_2(t) = K_t$. In
a Lagrangian perspective, we can interpret~$\theta_t$ as a momentum,
and thus the matrix~$K_t$ as an acceleration field.
Moreover, in dimension~$d=1$, we can show going one order further in the computations of
Appendix~\ref{sec:proofhigher} that the third term $T_3(t) = Q_t$ is the solution to:
\[
  \left\{\begin{aligned}
&  \dot{Q}_t +  b'''\theta_t + 3 b''K_t + 3b'Q_t+6 K_t D Q_t   = 0,
  \\& Q_T   =  f'''(\phi_T),
  \end{aligned}
  \right.
\]
where the derivatives of~$b$ are evaluated at the instanton~$\phi_t$.
The next terms follow similarly by computing the next orders of the Taylor expansion.

We finally note that similar computations appear in~\cite{bouchet2016perturbative}. However,
the setting
of this paper is different since the authors consider a drift~$b^\lambda$ depending on
a free parameter~$\lambda$ unrelated to the temperature, and the expansions are realized
with respect to this additional degree of freedom. To the best of our knowledge, expanding
around the noiseless characteristic equation is a new technique.
\end{remark}

%%%%%%%%%%%%%%%%%%%%%%%%%%%%%%%%%%%%%%%%%%%%%%%%%%%%%%%%%%%%%%%%%%%%%%%%%%%%%%%%%%%%%%%%

\subsection{Perturbative formula for the free energy at finite temperature}
\label{sec:Zexpansion}

In the previous section, we focused on constructing an approximate
optimal control for Monte Carlo importance sampling estimators. In
addition to this result, we now deduce from~\eqref{eq:gexpand} a
perturbative formula for~$(Z_\eps)_{\eps>0}$ for small values
of~$\eps$. For this, it suffices to note that
\[
Z_\eps = g_\eps(0, x_0).
\]
Considering~\eqref{eq:gexpand} for $t=0$ and $x=x_0$ then leads to
  \begin{equation}
    \label{eq:Zexpand}
    Z_\eps = Z^1_\eps(0)+ \mathrm{o}(\eps) = f(\phi_T) - \int_0^T \theta_t\cdot D\theta_t\,dt +
    \frac{\eps}{2}\int_0^T D:K_t\,dt\,.
 \end{equation}

  For brevity, we will denote~$Z^1_\eps(0)$ by~$Z^1_\eps$ in what follows. The above formula
  provides the first order correction to the zeroth-order
  Freidlin--Wentzell asymptotics~\eqref{eq:Z0}. As mentioned in
  Remark~\ref{rem:higher}, we could continue to construct higher order
  corrections and obtain a full expansion of the free energy at finite
  temperature through integrals of solutions to ordinary differential
  equations. Since we are more interested on the numerical side in
  this paper, we propose~\eqref{eq:Zexpand} as a way to numerically
  correct the Freidlin--Wentzell asymptotics~\eqref{eq:Z0} for small
  temperatures without resorting to Monte Carlo simulation.

\begin{remark}[Relation to prefactor analysis]
  \label{rem:Zeps}
  The correction term in~\eqref{eq:Zexpand}
  reads
  \[
  \int_0^T \mathrm{Tr}(D K_t)\,dt,
  \]
  meaning that the correction to~$A_\eps$ defined in~\eqref{eq:Aeps} is
  \[
  \exp\left(   \frac{1}{2}\int_0^T \mathrm{Tr}(D K_t)\,dt \right).
  \]
  Defining a matrix $G\in\R^{d\times d}$ via
  \begin{equation*}
    \dot G_t = G_t D K_t\,,
  \end{equation*}
  we can apply Liouville's formula\footnote{For a given matrix-valued process
    $A:[0,T]\to\R^{d\times d}$, the matrix $\Psi(t):[0,T]\to\R^{d\times d}$
    solution to $\dot\Psi(t) =A(t)\Psi(t)$
    satisfies $\det\Psi(T) = \det \Psi(0) \exp\left(\int_0^T\mathrm{Tr} A(t)\,dt\right)$.} to obtain
    \[
    \exp\left(   \frac{1}{2}\int_0^T \mathrm{Tr}(D K_t)\,dt \right)
    = \frac{\sqrt{\mathrm{det}\,| G_T|}}{\sqrt{\mathrm{det}\, |G_0|}}.
    \]
    It is common to express the first correction to the
    Freidlin--Wentzell small temperature asymptotics as the
    determinant of a Hessian matrix. As a result, the perturbative
    formula~\eqref{eq:Zexpand} can be understood as a prefactor
    analysis, and the integral of the Riccati matrix as a continuous
    version of the determinant prefactor that arises for instance in
    the Eyring--Kramers formula (see
    \textit{e.g.}~\cite{bouchet2016generalisation} and references
    therein, as well as~\cite{grafke-schaefer-vanden-eijnden:2021,
      schorlepp-grafke-grauer:2021}). Following
    Remark~\ref{rem:higher}, our methodology allows to compute $Z_\eps
    = Z^M_\eps + \mathrm{o}(\eps^M)$ where~$Z^M$ is a power series
    in~$\eps$ up to order~$M\in\mathbb{N}$, with coefficients defined
    as integrals of solutions to ordinary differential equations.
\end{remark}

\begin{remark}[Multiple instantons]
  \label{rem:multiple}
  For now it is clear that our strategy relies on the well-definedness of the reaction path.
  As we said, this is nothing else than the characteristic solution to
  the noiseless HJB equation~\eqref{eq:HJBzero} associated with the optimal control.
  However, in many cases the characteristic is
  ill-defined, which provokes shocks and discontinuities in solutions to the HJB problem.
  This is why a theory of weak solutions has been developed, in order to provide a sense
  of solution in cases where a classical solution does not exist.

  In these more complicated (yet easy to construct)
  situations~\cite{glasserman-wang:1997, vanden-eijnden-weare:2012},
  it is not clear yet how to adapt our method. Depending on the
  problem one wishes to solve, it may be possible to content oneself
  with the ``most important'' instanton, that is the one defining the
  Freidlin--Wentzell asymptotics. Otherwise, one may want to consider
  several characteristics and glue their resulting expansions together
  appropriately. We will not address this issue here, and thus our
  results, as shown, only apply to the situation where the
  characteristic system is well-defined. Understanding how our
  methodology can be extended to situations where only a weak solution
  is available is an interesting open problem.
\end{remark}

\begin{remark}[Error analysis]
  Provided the problems raised in the above remark are addressed properly,
  controlling precisely the error terms in~\eqref{eq:gexpand} is another interesting
  mathematical problem. We believe this can be tackled by more standard error analysis
  techniques~\cite{freidlin1998random,lelievre2016partial}. However, even if such
  error estimates were
  available, it is yet another problem to prove that the resulting importance sampling
  estimator built on~\eqref{eq:g2} indeed reduces the variance for
  estimating~\eqref{eq:Aeps}. This is a subtle problem for which we refer
  to~\cite{sadowsky1990large,glasserman-wang:1997,asmussen2007stochastic,guyader2020efficient}
  and references therein.
\end{remark}

%%%%%%%%%%%%%%%%%%%%%%%%%%%%%%%%%%%%%%%%%%%%%%%%%%%%%%%%%%%%%%%%%%%%%%%%%%%%%%%%%%%%

\section{Numerical applications}
\label{sec:lowtemp:applications}

In the following section, we demonstrate the usefulness of our
approximation by performing numerical experiments on a number of
example systems, comparing first the value of the free energy~$Z_\eps$
estimated by Monte-Carlo sampling to the Freidlin--Wentzell
asymptotics~$Z^0$ (which is constant) and the linear
approximation~$Z^1_\eps$.  Further, we compare the performance of a
naive (unbiased) Monte-Carlo estimator to the one using importance
sampling with the approximate optimal biases~$g^0$ and $g^1$.

For comparing Monte Carlo estimators, we use the \emph{relative error},
which is  the ratio
of the standard deviation of our estimator over its average for a number
of realizations. If indeed
our approximation to the optimal bias is effective, heuristically we
expect smaller relative error for higher order approximations to the
optimal bias. Numerically measuring the relative error is therefore  an
experimental quantification of the variance reduction capabilities of
our proposed estimators. Moreover for all the numerical simulations we discretize
the underlying SDE with a standard Euler--Maruyama scheme with time
step~$\Delta t >0$, and neglect the error arising from this
numerical quadrature~\cite{kloeden1992}.

\subsection{One dimensional Ornstein--Uhlenbeck process}

The simplest situation is the one dimensional case where the drift is given by
$b(x) =-\gamma x$, with $\gamma>0$. We further set $\sigma=1$
and $f(x) = x$. This particular case corresponds to the
Ornstein--Uhlenbeck process.

For the above choice, we know that the optimal control is actually
equal to the conjugate momentum~$(\theta_t)_{t\in[0,T]}$ from
equation~(\ref{eq:instanton}). Therefore, the zeroth order
approximation~\eqref{eq:g1} actually already provides the zero
variance estimator described in Section~\ref{sec:IBP}. Numerically, we
therefore expect an estimator with variance close to zero. The first
order correction term should not improve the results, so the matrix
(scalar in this case)~$(K_t)_{t\in[0,T]}$ should be zero for all times
here.

The numerical experiment is performed with $x_0=-1$, $T=10$ and
$\Delta t=0.01$ for the numerical discretization, performing $N=10^6$
experiments. The results are shown in
Table~\ref{tbl:OU}, where we compare the naive unbiased estimator to
the estimator biased with the instanton (zeroth order approximation)
and the estimator biased to first order. While the relative error of
the naive estimator blows up with decreasing~$\eps$, the relative
error is zero in both the zeroth and first order estimators, implying
that the first order estimator is already equivalent to the optimal
zero variance bias. We also numerically observe that the Riccati
matrix is indeed equal to zero (not shown).

\begin{table}[h]
  \begin{center}
    \begin{tabular}{rrrr}
      \hline\hline
      $\eps$ & Naive estimator & Zeroth order estimator & First order estimator \\
      \hline
      1 & 0.81& 2.73$\cdot 10^{-12}$& 3.15$\cdot 10^{-12}$\\
      0.5 & 1.31& 6.12$\cdot 10^{-12}$& 1.05$\cdot 10^{-11}$\\
      0.1 & 9.28& 5.04$\cdot 10^{-12}$& 3.08$\cdot 10^{-13}$\\
      0.05 & 40.38& 1.03$\cdot 10^{-11}$& 3.74$\cdot 10^{-12}$\\
      0.01 & 291.82& 8.71$\cdot 10^{-12}$& 1.2$\cdot 10^{-12}$\\
  \hline
\end{tabular}
  \end{center}
  \caption{Relative error for $Z_\eps$ for the Ornstein--Uhlenbeck test case,
    comparing the naive Monte Carlo estimator with the zeroth and
    first order biased Monte Carlo estimators. \label{tbl:OU}}
\end{table}

\subsection{Two-dimensional nonlinear nonequilibrium process}

We now break detailed balance by considering a drift~$b(x)$ that is
not the gradient of a potential~$V(x)$. In that case it is no longer
true that the reaction path is merely a reverse relaxation trajectory
driven by the potential level sets, and the reaction path itself must
be computed by numerically solving the instanton
equations~\eqref{eq:instanton}, which is a well-established problem in
the literature~\cite{e-ren-vanden-eijnden:2004,
heymann-vanden-eijnden:2008, grafke-grauer-schaefer:2015,
grafke-schaefer-vanden-eijnden:2017}. Here, we use the algorithm
from~\cite[Section III.A]{grafke-vanden-eijnden:2019}. Moreover, the
optimal control is also no longer explicit like in the
Ornstein--Uhlenbeck case.

As an example, we take the system 
\begin{equation}
  \label{eq:2d-nonlinear}
  \begin{cases}
    dX_t = (Y_t^3 - X_t^3)\,dt + \sqrt{\eps}\,dB_t^X\\
    dY_t = (-X_t^3 - Y_t^3)\,dt + \sqrt{\eps}\,dB_t^Y,
  \end{cases}
\end{equation}
for two independent Brownian motions~$(B_t^X,B_t^Y)_{t\geq 0}$, The
dynamics experiences a nonlinear attractive force towards the unique
fixed point $(x,y) = (0,0)$ with a nonlinear swirl in clockwise
direction that becomes stronger away from the origin. As further
complication, we choose a finite time interval~$T$ for the transition
to happen, and start away from the fixed point.

As observable in~(\ref{eq:Aeps}) we take $f(x,y) = x$, biasing the
dynamics towards large values of the $x$-component of the process. We
start at $(x_0,y_0)=(-1,-1)$, away from the fixed point, and run the
process for $T=10$, which is long enough so that the instanton is not
a straight line, but short enough so that it does not completely relax
to the fixed point and then leaves it again at a later time (as would
be the case in the limit $T\to\infty$). The resulting event is
therefore a complicated interplay between the nonlinear dynamics and
the conditioning on large $x$ values, and the expected distribution of
end-points is far from the invariant measure.

\begin{table}[h]
  \begin{center}
    \begin{tabular}{rrrr}
      \hline\hline
      $\eps$ & Naive estimator & Zeroth order estimator & First order estimator \\
      \hline
      0.5 & 0.04 &   0.02 &  0.113\\
      0.2 & 0.089 &  0.029 & 0.135\\
      0.1 & 0.253 &  0.036 & 0.09 \\
      0.05 & 1.005 &  0.048 & 0.13 \\
      0.02 & 6.793 &  0.094 & 0.133\\
      0.01 & 15.592 &   0.753 &  0.155\\
      0.005 & 25.709 &   0.631 &  0.136\\
      \hline
 \end{tabular}
  \end{center}
  \caption{Relative error for $Z_\eps$ for the two-dimensional test problem,
    comparing the naive Monte Carlo estimator with the zeroth and
    first order biased Monte Carlo estimators.\label{tbl:2D}}
\end{table}

The numerical parameters are $\Delta t=10^{-2}$ and $N=10^6$
experiments. The results are shown in Table~\ref{tbl:2D}, where it can
be seen that for $\eps\to0$ the relative error blows up for the naive
estimator, while it is roughly constant for the zeroth and first order
estimators, the later being smaller.

\begin{figure}[h]
  \includegraphics[width=\textwidth]{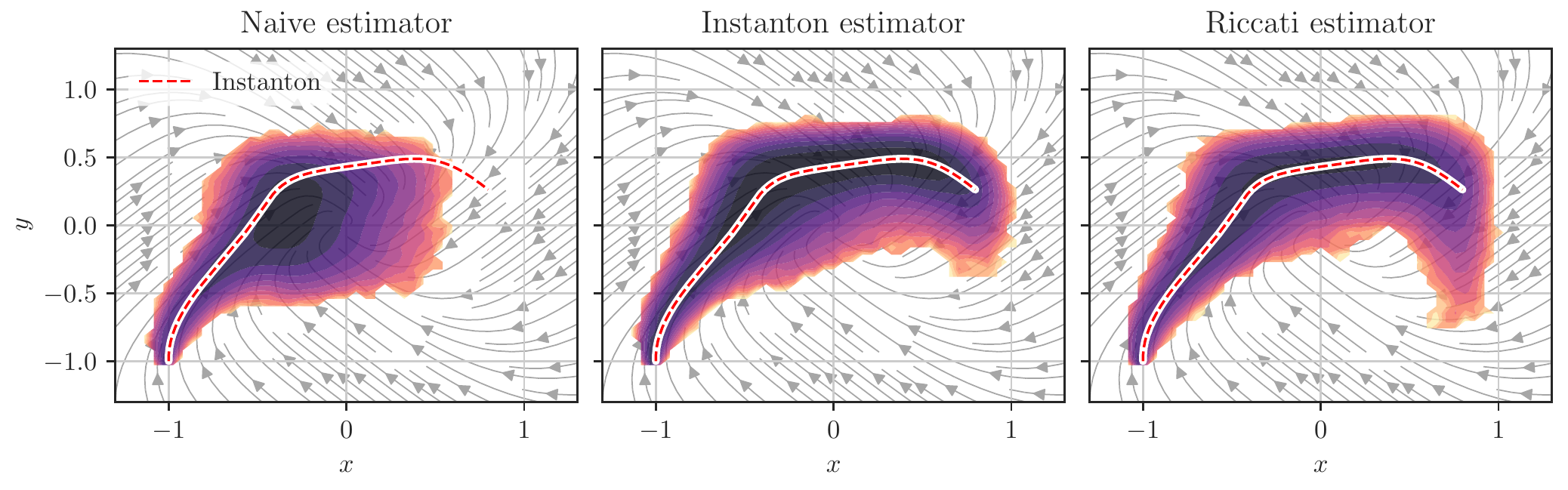}
  \caption{Nonlinear two-dimensional process with swirl as defined in
    equation~\eqref{eq:2d-nonlinear} represented by arrows. The
    red dashed line depicts the
    instanton. The heat map is a histogram of the trajectories of the
     Monte-Carlo experiment for $\eps=0.01$. For the naive estimator
    (left), most sample trajectories follow the deterministic decay
    trajectory by gathering around the fixed point~$(0,0)$, so that
    large values for~$x$ are rarely observed. For the zeroth order
    control (center) and the first order one (right) the sample
    trajectories are staying more closely around the instanton, but to
    a different degree at different locations.\label{fig:2D}}
\end{figure}

\begin{figure}[h]
  \includegraphics[width=0.48\textwidth]{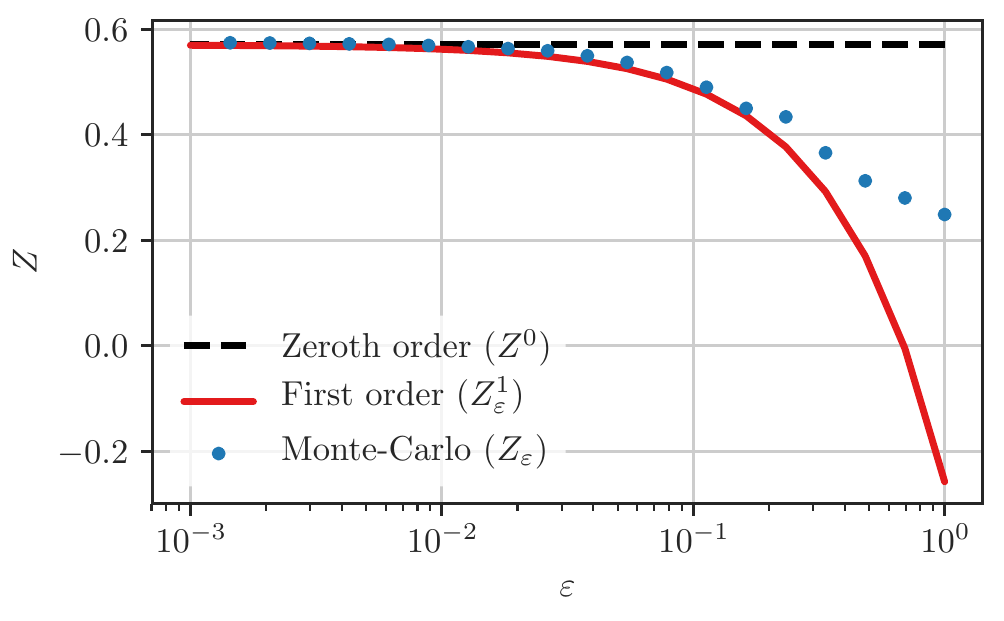} \includegraphics[width=0.48\textwidth]{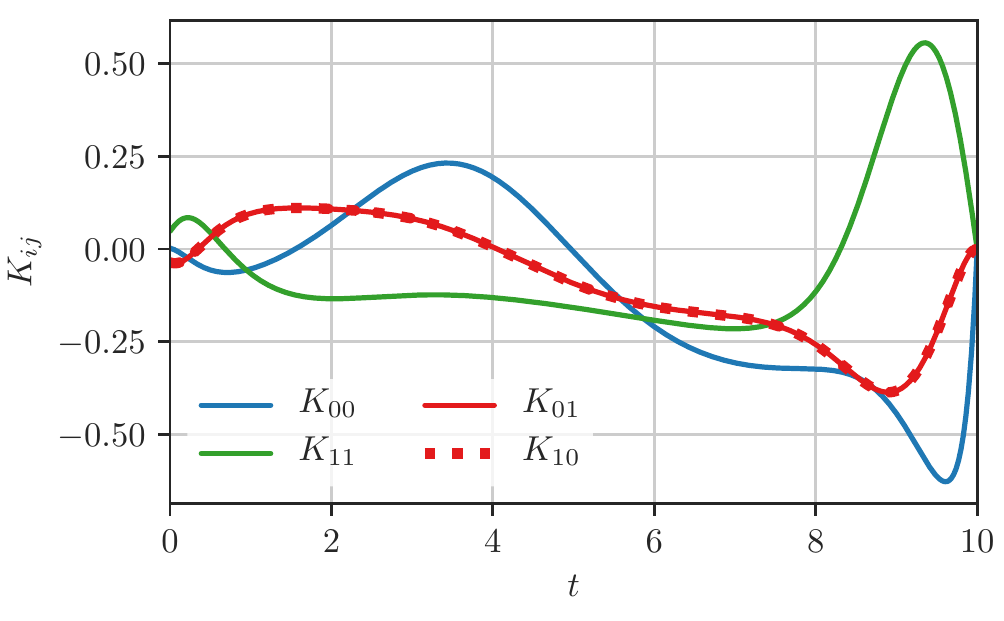} \caption{Left:
  Validity of the approximation of~$Z_\eps$ in the two-dimensional
  problem. In the limit of small~$\eps$, the Monte Carlo estimator
  agrees with the constant zeroth order~\eqref{eq:Z0}. For larger
  values of~$\eps$, the values of~$Z_\eps$ depart from the
  constant~$Z^0$. The first order approximation~$Z_\eps^1$ captures
  this departure for at least an order of magnitude in~$\eps$. For still larger
  values,~$Z_\varepsilon^1$ and~$Z_\eps$ diverge as expected. Right:
  Evolution of the four components of~$K_t$ in the
  two-dimensional problem.\label{fig:2D_Z_K}}
\end{figure}

Figure~\ref{fig:2D} compares the different sampling procedures for
$\eps=0.01$. In the naive estimator, most trajectories cluster around
the deterministic decay path, swirling in clockwise direction into the
origin, and consequently not reaching a large value of~$x$. With the
zeroth order estimator, and to a different degree with the first order
one, the samples remain closer to the instanton (with different
strength in different regions). Figure~\ref{fig:2D_Z_K} (left)
shows~$Z_\eps$,~$Z^0$ and~$Z^1_\eps$ as a function of~$\eps$.
While~$Z^0$ captures the constant,
$\eps$-independent limiting value of~$Z_\eps$, the departure
of~$Z_\eps$ from this constant is captured accurately by the first
order approximation~$Z_\eps^1$ for a prolonged interval in~$\eps$. As
expected, for larger values of~$\eps$, higher order effects come into
play, degrading the accuracy of the expansion, which could be improved
by considering higher order terms (see Remark~\ref{rem:Zeps}).
Finally, Figure~\ref{fig:2D_Z_K} (right) shows the evolution of the
$2\times2$ matrix~$(K_t)_{t\in[0,T]}$ along the instanton trajectory
which is used to compute the approximate optimal bias
via~\eqref{eq:g2}-\eqref{eq:EDOK}.

\subsection{Double well potential}

We next consider a one dimensional double-well potential with
$V(x) = \frac14(x^2-1)^2$, which has locally stable fixed points at $x=\pm 1$, and
set $b(x) = - \nabla V(x)$ and $f(x)=x$. We are
starting the process in the left fixed point  $x_0=-1$, so that a
typical fluctuation leading to high values of~$f$ corresponds
to a trajectory crossing to the right well, which becomes a rare event
in the low~$\eps$ limit.

This example is more complicated than the previous ones because two
fixed points exist.  As a consequence, in the non-convex regions of the
potential,
straying from the globally optimal path is amplified by the dynamics
because forward trajectories are spreading. Interestingly, this
problem is more or less pronounced depending on the given time
interval~$T$: for shorter transition times, the kinetic
term~$\dot\phi^2$ in the Freidlin--Wentzell action dominates, and the
dynamics become comparably unimportant. In order to illustrate
this phenomenon in our numerical
experiments, we choose several values for the final time~$T$.

\begin{figure}[h]
  \includegraphics[width=\textwidth]{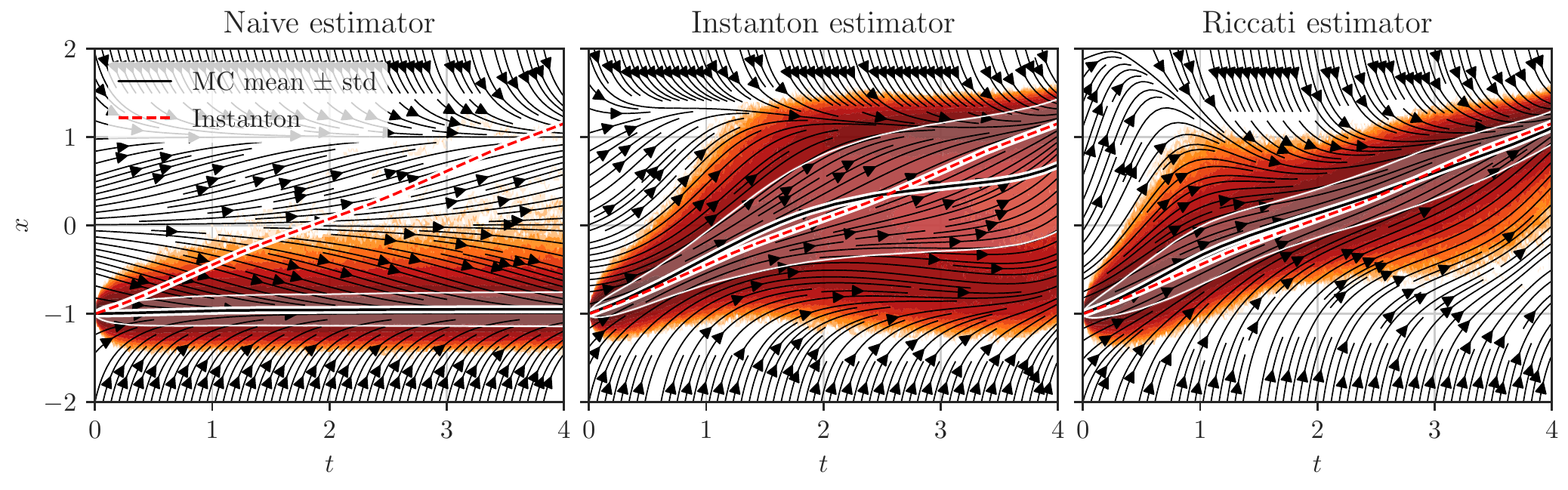}
  \caption{One-dimensional double-well process for $\eps=0.01$. The
    red dashed line depicts the instanton, while the black solid line
    and surrounding gray shading displays the (first order biased)
    Monte-Carlo estimator mean and one standard deviation region
    around it. The heat map is a histogram of all trajectories. The
    black streamlines display the total drift field, \textit{i.e.}~the
    sum of the drift~$b(x)$ and the bias respectively equal to
    zero,~$D\nabla g^0(t,x)$ and~$D\nabla g^1(t,x)$ in the plots from
    left to right. For the naive estimator (left), almost no sample
    trajectory leaves the lower basin, leading to a bad estimate of
    the expectation. For the instanton estimator (center), many sample
    trajectories transition to the upper basin, with a wide
    variance. In the Riccati first order estimator (right),
    trajectories are staying more closely around the instanton
    trajectory, which is the optimal one in the small temperature
    limit.\label{fig:1DDouble}}
\end{figure}

Again, we compute the instanton via the algorithm from~\cite[Section
  III.A]{grafke-vanden-eijnden:2019}. The sampling procedure, where
numerical parameters are set to $\Delta t=10^{-2}$, $T=4$, and
$N=10^6$, is depicte in Figure~\ref{fig:1DDouble}. The heat maps
display histograms of the sample trajectories. While for the naive
estimator only very few manage to transition to the upper basin, many
more are driven across the barrier with the instanton drift active. In
the first order case, the trajectories are kept in a tube around the
instanton, so that a majority of trajectories explore the space around
the optimal trajectory at small temperature. We note however that the
force field has a surprising behavior far from the instanton, which we
expect given the estimate~\eqref{eq:gexpand}. The corresponding
relative errors are listed in Table~\ref{tbl:1DDouble}. In particular,
while the relative error explodes with $\eps\to0$ for the naive
estimator, it remains roughly constant for the zeroth order estimator and
decreases significantly for the first order one. For $T=8$ (all other
parameters being the same), the observation is quite different. In
fact, as shown in Table~\ref{tbl:1DDouble8}, for this time and even
longer ones, variance reduction is no longer clearly obtained. This
illustrates that the performance of our approximation of the optimal
bias obtained here, which is reached under strong assumptions
(uniqueness of the instanton in particular, see the beginning of
Section~\ref{sec:gexpansion} and Remark~\ref{rem:multiple}), may be deteriorated in
non-convex cases for a large final time~$T$.

\begin{figure}[h]
  \begin{center}
    \includegraphics[width=0.8\textwidth]{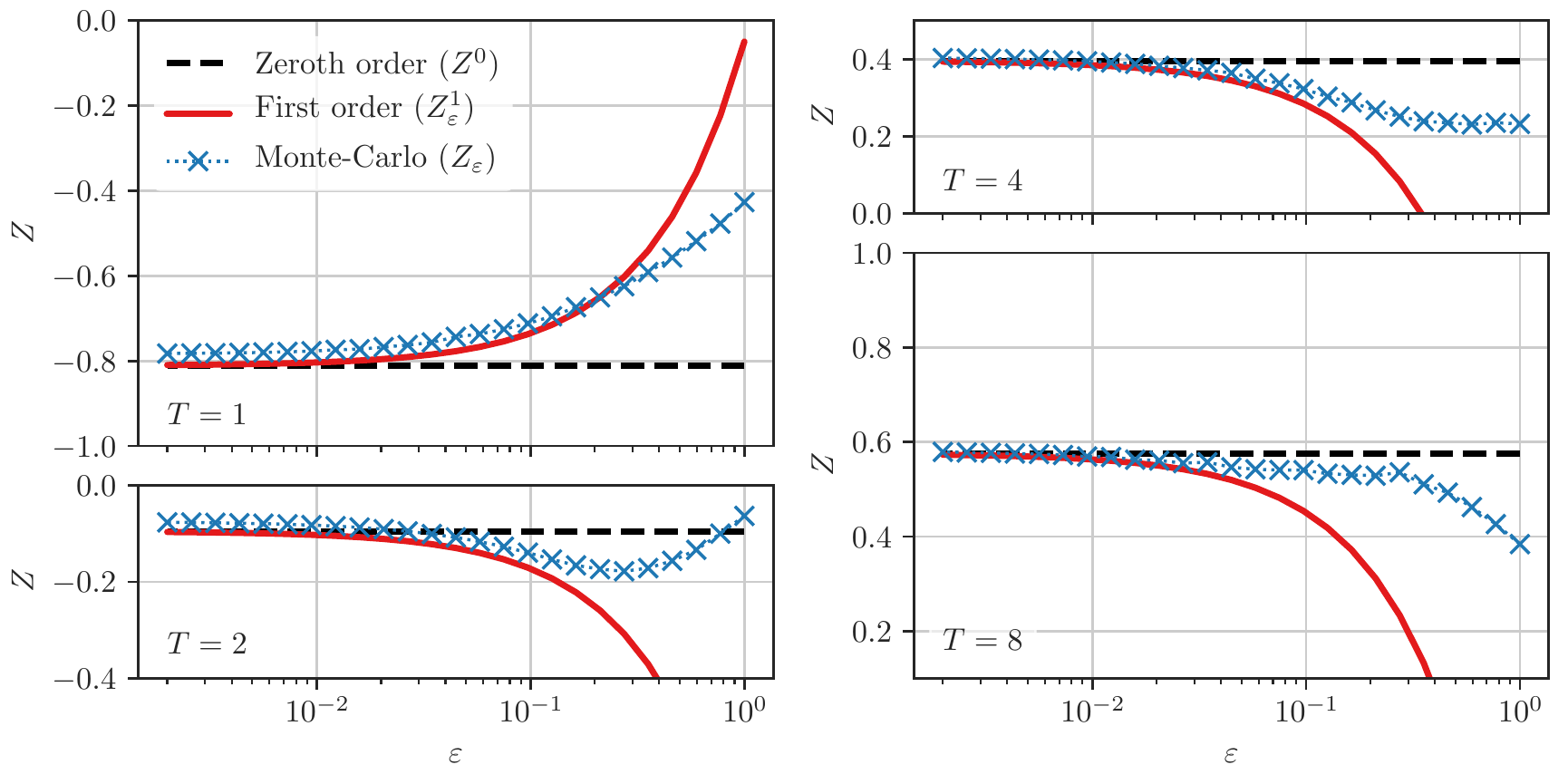}
  \end{center}
  \caption{One-dimensional double-well process, validity of the
    approximation of the free energy~$Z_\eps$.
    For $\eps\to0$, the constant approximation
    predicts the right limiting value~$Z^0$. Departure from this constant
    value is correctly approximated by the first order
    approximation~$Z_\varepsilon^1$. The Monte Carlo estimated values are obtained
    using trajectories biased to first order as above.\label{fig:doublewell-Zeps}}
\end{figure}

However, the fact that the relative error ceases to decrease for $\eps\to0$ in
the first order approximation does not necessarily mean that the
approximation~$Z_\eps^1$ fails as
well. In fact, as shown in Figure~\ref{fig:doublewell-Zeps},~$Z_\eps$
is well-approximated by~$Z^0$ and~$Z_\eps^1$
up to $T=8$. The constant value of the zeroth order
term correctly approximates the limiting value of the numerical
experiment, and the departure from that constant is correctly captured
by the first order correction.

\begin{table}[h]
  \begin{center}
    \begin{tabular}{rrrr}
      \hline\hline
      $\eps$ & Naive estimator & Zeroth order estimator & First order estimator \\
      \hline
      1 & 0.96& 1.34& 13.6\\
      0.5 & 2.11& 2.55& 14.8\\
      0.1 & 81.21& 8.98& 25.3\\
      0.05 & 800.99& 6.57& 5.58\\
      0.01 & 343.97& 7.21& 1.71\\
      0.005 & 835.89& 5.26& 0.456\\
    \hline
\end{tabular}
  \end{center}
  \caption{Relative error for $Z_\eps$ for the double-well test case,
    $T=4$, comparing the naive Monte Carlo estimator with the zeroth
    and first order biased Monte Carlo estimators.\label{tbl:1DDouble}}
\end{table}

\begin{table}[h]
  \begin{center}
    \begin{tabular}{rrrr}
      \hline\hline
      $\eps$ & Naive estimator &  Zeroth Order estimator & First order estimator \\
      \hline
      1 & 0.89& 1.53& 67.9\\
      0.5 & 1.65& 3.14& 50.1\\
      0.1 & 32.66& 167.0& 47.1\\
      0.05 & 656.07& 42.0& 177.0\\
      0.01 & 541.15& 144.& 15.6\\
      0.005 & 721.8& 29.1& 33.1\\
    \hline
\end{tabular}
  \end{center}
  \caption{Relative error for the double-well test case, $T=8$,
    comparing the naive Monte Carlo estimator with the zeroth and
    first order biased Monte Carlo estimators.\label{tbl:1DDouble8}}
\end{table}

%%%%%%%%%%%%%%%%%%%%%%%%%%%%%%%%%%%%%%%%%%%%%%%%%%%%%%%%%%%%%%%%%%
\section{Discussion}
\label{sec:discussion}

In this paper, we studied the task of computing a free energy-like
quantity as it commonly arises in statistical physics. For this, we consider the
optimal control problem associated with finding the optimal bias to
reduce the variance of an importance sampling Monte Carlo estimator.
We propose a new methodology to approximate the solution to the
optimal stochastic control problem. From a physics perspective, it
corresponds to designing a non-homogeneous Taylor expansion around the
instanton of the dynamics. From a partial differential equation
standpoint, it is an expansion around the characteristic curve of the
zero-temperature limit of the Hamilton--Jacobi--Bellman equation associated with
the optimal control.  Our
approach differs from the more standard technique of expanding the
solution in the temperature parameter (sometimes called WKB ansatz) since
here the temperature appears only implicitly through the distance to
the instanton of typical reacting trajectories, and the expansion
is defined through solutions to ordinary differential equations
instead of partial differential equations. 

With this new tool at our disposal, we achieve two goals. First, we
use our approximation to reduce the variance with a standard
tilting procedure, replacing the optimal control with our expansion
computed  offline from simple ordinary differential
equations. This approach therefore yields a very efficient method to
approximate the optimal control close to the most likely realization
given by the instanton, which is exactly the right regime in the low
temperature limit. Next, we derive a new formula for expanding the
free energy in the small temperature parameter, which refines the
Freidlin--Wentzell asymptotics. We explicitly compute the first order
term of the series and explain how to pursue the expansion to any order.

Finally, we propose a series of examples to illustrate the validity of
our methodology. We demonstrate how the suggested approximate optimal
control reduces variance in a set of numerical examples, and how the
free energy expansion extends the Freidlin--Wentzell asymptotics. We
further show some limitations of our approach, in particular concerning
variance reduction in a non-convex setup and for large final times.

As one can note, the theoretical arguments we use are based on quite
stringent conditions: well-posedness and smoothness of the solution to
the Hamilton--Jacobi--Bellman problem, uniqueness of the instanton,
various boundedness assumptions hidden in the expansion analysis,
etc. This raises a number of questions on how in its current state the
approach may fail, for instance when several solutions exist to the
characteristic equation and the HJB equation only has a weak, non-smooth
solution. We know that this situation naturally arises in many contexts,
which can be interpreted as a non-convexity of the rate function from
a large deviations perspective, the creation of shocks from a partial
differential equation point of view, or caustics from the physicist's
viewpoint. On the other hand, from a computational perspective, even
though we could quantify the closeness of the expansion to the optimal
control, it is yet difficult to assess that the tilted estimator
indeed reduces the variance in simulations in
general~\cite{sadowsky1990large, glasserman-wang:1997,
asmussen2007stochastic,guyader2020efficient}. Our proposed control is
therefore only the first step in the direction of rigorously
establishing how and when the optimal control can be expanded around
the instanton, improving on existing suggestions to use the instanton
as an approximation for the optimal tilt as a mere heuristic in
importance sampling, for instance in cloning
algorithms~\cite{wouters-bouchet:2016, grafke-vanden-eijnden:2019} or
in instanton biased importance sampling motivated from path integral
techniques~\cite{ebener-margazoglou-friedrich-etal:2019}. Our results
suggest that the approximation is built in such a way that variance is
indeed reduced in the small temperature limit for simple systems, but
this remains to be proved rigorously, and under which precise
conditions.  Although studying such issues possibly requires a
significant effort, we hope the possible applications both in
numerical and theoretical directions will motivate further research in
this direction.

%%%%%%%%%%%%%%%%%%%%%%%%%%%%%%%%%%%%%%%%%%%%%%%%%%%%%%%%%%%%%%%%%%%%%%%%%%%%%%%%%%%%%%%
\section*{Acknowledgments}
%%%%%%%%%%%%%%%%%%%%%%%%%%%%%%%%%%%%%%%%%%%%%%%%%%%%%%%%%%%%%%%%%%%%%%

The authors warmfully thank Eric Vanden-Eijnden for his insightful
advice on the work, as well as Gabriel Stoltz and Hugo Touchette for
interesting discussions. The authors are also grateful towards the referees
for providing valuable input and mentioning interesting references.
The PhD of Grégoire Ferré was supported by
the Labex Bézout ANR-10-LABX-58-01 and a grant from \'{E}cole des
Ponts ParisTech.  Grégoire Ferré is grateful to the CERMICS laboratory
for funding a two-month stay at the Courant Institute of Mathematical
Sciences in New York where the project was initiated. Tobias Grafke
acknowledges the support received from the EPSRC projects EP/T011866/1
and EP/V013319/1.

\appendix
\section*{Appendix}

\section{Proofs of Section~\ref{sec:IBP}}
\label{sec:proofscontrol}

We first prove an integration by part formula by showing that, given the
dynamics~\eqref{eq:X} and~\eqref{eq:Xtilde}, the
expectation in~\eqref{eq:Aeps}  rewrites
  \begin{equation}
    \label{eq:IBP}
    A_{\eps} =\e^{g(0, x_0)} \E_{x_0} \left[ \exp\left(\frac{1}{\eps}
        \big[f(\widetilde X_T^{\eps}) - g(T,\widetilde X_T^{\eps}) \big]
        + \frac{1}{\eps} \int_0^T \alpha(t,\widetilde X_t^{\eps})\,dt
      \right)\right],
  \end{equation}
  where
  \begin{equation}
    \label{eq:alpha}
    \alpha(t,x) =\partial_t g(t,x) + \LL g(t,x) + \frac{1}{2} |\sigma\nabla g|^2(t,x). 
  \end{equation}

The proof of this formula relies on the Girsanov theorem and the gradient structure of the drift.
  We first write the Girsanov formula for the path change of
  measure~\cite{karatzas2012brownian,bellet2006ergodic} between the
  processes~$(X_t^{\eps})_{t\geq 0}$ and~$(\widetilde X_t^{\eps})_{t\geq 0}$
  (provided technical conditions are met):
  \begin{equation}
    \label{eq:intgirs}
    A_{\eps}=\E_{x_0} \left[ \e^{\frac{1}{\eps} f(X_T^{\eps})}\right]
    = \E_{x_0} \left[ \e^{\frac{1}{\eps} f(\widetilde X_T^{\eps})
        - \frac{1}{2\eps}
        \int_0^T |\sigma \nabla g|^2(t,\widetilde X_t^{\eps})\, dt - \frac{1}{\sqrt{\eps}} \int_0^T
        \sigma \nabla g(t,\widetilde X_t^{\eps})\, dB_t   }\right].
  \end{equation}
  We now use It\^o formula over a trajectory of~$(\widetilde X_t^{\eps})_{t\geq 0}$ using the
  generator~\eqref{eq:gen}:
  \[
  d g(t,\widetilde X_t^{\eps}) = \big( \partial_t g + \LL g
  + \nabla g\cdot D  \nabla g\big)(t,\widetilde X_t^{\eps})\,dt
  +\sqrt{\eps} \sigma \nabla g(t,\widetilde X_t^{\eps}) \,dB_t.
  \]
  Integrating in time and dividing by~$\eps$, the above equation becomes
  \[
  -\frac{1}{\sqrt{\eps}} \int_0^T\sigma \nabla g(t,\widetilde X_t^{\eps}) \,dB_t
  =  - \frac{g(T,\widetilde X_T^{\eps}) - g(0,\widetilde X_0)}{\eps}
  + \frac{1}{\eps} \int_0^T \big( \partial_t g + \LL g + |\sigma \nabla g|^2\big)
  (t,\widetilde X_t^{\eps})\,dt.
  \]
  Inserting this equality into~\eqref{eq:intgirs} leads to~\eqref{eq:IBP}.

%%%%%%%%%%%%%%%%%%%%%%%%%%%%%%%%%%%%%%%%%%%%%%%%%%%%%%%%%%%%%%%%%%%%%%%%%

We next turn to the derivation of the optimal control~\eqref{eq:gopt}.
We first note, using the Feynman--Kac formula~\cite[Theorem~21.1]{kallenberg2006foundations},
  that~$\psi_{\eps}$ is the solution to the following backward PDE:
  \begin{equation}
    \label{eq:feynmankacpsiA}
    \left\{
    \begin{aligned}
      \partial_t\psi_{\eps}  + \LL\psi_{\eps} & = 0 & \\
      \psi_{\eps}(T,x) &= \e^{\frac{1}{\eps}f(x)}, & \forall\,x\in\R^d.
    \end{aligned}
    \right.
    \end{equation}  
  Defining $g_{\eps} =\eps \log \psi_{\eps}$, we see
  that the time derivative of~$g_{\eps}$ reads
  \[
  \begin{aligned}
    \partial_t g_{\eps} & =\eps \frac{\partial_t\psi_{\eps}}{\psi_{\eps}}
    =\eps \frac{-\LL \psi_{\eps} }{\psi_{\eps}} 
    \\ & = -\eps\e^{-g_{\eps}/\eps}\LL \e^{g_{\eps}/\eps} =\eps \Big(
    -\eps^{-1} b\cdot \nabla g_{\eps}
    - \e^{-g_{\eps}/\eps}  \frac{\sigma\sigma^\mathrm{T}}{2}:\big(\nabla
    (\e^{g_{\eps}/\eps}\nabla g_{\eps} ) 
    \big)\Big)
    \\ & =\Big( -\LL g_{\eps} - \frac{1 }{2} \left| \sigma\nabla g_{\eps}
    \right|^2\Big).
  \end{aligned}
  \]
  Using the terminal condition in~\eqref{eq:feynmankacpsiA} shows that~$g_\eps$ is
  the solution to
  \begin{equation}
    \label{eq:feynmankacphiA}
\left\{
    \begin{aligned}
      \partial_t g_{\eps}  + \LL g_{\eps}
      + \frac{1}{2} \left| \sigma\nabla g_{\eps} \right|^2  &= 0 & \\
      g_{\eps}(T,x) & = f(x) , & \forall\,x\in\R^d.
    \end{aligned}
    \right.
  \end{equation}
  As a result,~\eqref{eq:feynmankacphiA} ensures that $\alpha(t,x) = 0$. Together
  with the terminal condition, this shows that~\eqref{eq:gopt} defines a zero-variance
  control since the estimator is deterministic.

%%%%%%%%%%%%%%%%%%%%%%%%%%%%%%%%%%%%%%%%%%%%%%%%%%%%%%%%%%%%%%%%%%%%%%%%%%%%%%
\section{Proof of~\eqref{eq:EDOK}-\eqref{eq:gexpand}}
\label{sec:proofhigher}

  The idea is to rewrite the Feynman--Kac mode~$\psi_{\eps}$ defined in~\eqref{eq:psiA}
  with the integration by part~\eqref{eq:IBP} presented in Appendix~\ref{sec:proofscontrol} in
  order to exhibit the leading behavior in~$\eps$. Consider
  the dynamics~$(\widetilde X_t^{\eps})_{t\in[0,T]}$ defined in~\eqref{eq:Xtilde} with~$g^1$
  given by~\eqref{eq:g2}. Using the Girsanov theorem like in Appendix~\ref{sec:proofscontrol},
  starting from any $t\geq 0$ and $x\in\R^d$, we have
   \begin{equation}
    \label{eq:psiepsg}
    \begin{aligned}
  \psi_{\eps}(t,x)  = & \E_{t,x}\left[
    \e^{ \frac{1}{\eps}( f(\widetilde X_T^{\eps}) - g^1(T,\widetilde X_T^{\eps}) + g^1(t,\widetilde X_t^{\eps}))
      +\frac{1}{\eps}\int_t^T \alpha(s,\widetilde X_s^\eps)\,ds}
    \right] \\
   =& \e^{\frac{g^1(t,x)}{\eps} } \E_{t,x}\left[
    \e^{ \frac{1}{\eps}( f(\widetilde X_T^{\eps}) - g^1(T,\widetilde X_T^{\eps}) )
      +\frac{1}{\eps}\int_t^T \alpha(s,\widetilde X_s^\eps)\,ds}
    \right],
  \end{aligned}
  \end{equation}
  where the function~$\alpha$ is defined in~\eqref{eq:alpha}. We now perform an expansion
  in~$\eps$ inside the expectation in~\eqref{eq:psiepsg}. First, the
  process~$(\widetilde X_t^{\eps})_{t\in[0,T]}$ admits the following expansion:
  \begin{equation}
    \label{eq:Xtildehigh}
  \widetilde X_t^{\eps} = \phi_t + \sqrt{\eps}\zeta_t + \eps\beta_t
  + \mathrm{O}\big(\eps^{3/2}\big),
  \end{equation}
  where
  \[\left\{
  \begin{aligned}
  d\zeta_t & = (\nabla b(\phi_t)\zeta_t + K_t\zeta_t)\,dt + \sigma\,dB_t,
  \\
  d \beta_t & = \big( \nabla b (\phi_t)\beta_t + \frac{1}{2} \zeta_t \cdot
  \nabla^2b(\phi_t) \zeta_t + K_t \beta_t\big) dt.
  \end{aligned}
  \right.
  \]
  This follows by expanding~$(\widetilde X_t^{\eps})_{t\in[0,T]}$ around the
  path~$(\phi_t)_{t\in[0,T]}$ and identifying
  the terms of different orders in~$\eps$ (by Taylor-expanding the drift~$b$).
  Note that we will actually not need the precise
  expression for the processes~$(\zeta_t)_{t\in[0,T]}$ and~$(\beta_t)_{t\in[0,T]}$ in what follows.

  We now come back to~\eqref{eq:psiepsg} by first considering the terminal terms.
  Using~\eqref{eq:Xtildehigh}, we obtain 
  \[
  \begin{aligned}
    f(\widetilde X_T^{\eps}) & = f(\phi_T) + \sqrt{\eps}\nabla f(\phi_T)\cdot \zeta_T
    + \eps\left[\nabla f(\phi_T)\cdot \beta_T + \frac{1}{2}\zeta_T\cdot
      \nabla^2 f(\phi_T) \zeta_T \right]+ \mathrm{O} \big(\eps^{3/2}\big)
    \\
    g^1(T,\widetilde X_T^{\eps}) & = \sqrt{\eps} \theta_T\cdot\zeta_T
    + \eps \left[ \theta_T\cdot \beta_T +
       \frac{1}{2}\zeta_T\cdot
      K_T \zeta_T \right]+ \mathrm{O} \big(\eps^{3/2}\big).
  \end{aligned}
  \]
  The terminal conditions for~$(\theta_t)_{t\in[0,T]}$ and~$(K_t)_{t\in[0,T]}$ lead to
  \begin{equation}
    \label{eq:hg}
    f(\widetilde X_T^{\eps}) - g^1(T,\widetilde X_T^{\eps}) = f(\phi_T)
    + \mathrm{O} \big(\eps^{3/2}\big).
  \end{equation}
  It remains to study the integral part in~\eqref{eq:psiepsg}, for which we
  expand the Girsanov weight~$\alpha$ with~\eqref{eq:Xtildehigh}. We have
  \begin{equation}
    \label{eq:alphahigh}
    \alpha(t,\widetilde X_t^{\eps}) = \partial_t g^1(t,\widetilde X_t^{\eps}) + b(\widetilde X_t^{\eps})\cdot \nabla g^1(t,\widetilde X_t^{\eps}) + \frac{\eps}{2}
    D:\nabla^2 g^1(t,\widetilde X_t^{\eps})
  + \frac{1}{2}|\sigma \nabla g^1(t,\widetilde X_t^{\eps})|^2.
  \end{equation}
  First, we notice that (we omit below the dependency of~$\phi_t$, $\theta_t$ and~$K_t$ on time for concision)
  \[
  \begin{aligned}
    \partial_t g^1(t,x) & = \dot{\theta}\cdot (x - \phi) - \theta\cdot \dot{\phi}
    + \frac{1}{2} (x - \phi)\cdot \dot{K} (x - \phi) - (x - \phi)\cdot K\dot{\phi}
    \\ \nabla g^1(t,x) & = \theta + K(x - \phi).
  \end{aligned}
  \]
  As a result,~\eqref{eq:alphahigh} reads
  \[
  \begin{aligned}
  \alpha(t,\widetilde X_t^{\eps}) =&\ \dot{\theta}\cdot (\widetilde X_t^{\eps} - \phi) - \theta
  \dot{\phi}  + \frac{1}{2} (\widetilde X_t^{\eps} - \phi)\cdot \dot{K} (\widetilde X_t^{\eps} - \phi)
   - \frac{1}{2}(\widetilde X_t^{\eps} - \phi)\cdot K\dot{\phi}
   - \frac{1}{2}(\widetilde X_t^{\eps} - \phi)\cdot K^\mathrm{T}\dot{\phi}
   + \frac{\eps}{2}
    D: K \\  &
  + b(\widetilde X_t^{\eps})\cdot\big( \theta + K(\widetilde X_t^{\eps} - \phi)\big)
  + \frac{1}{2}\big|\sigma\big(\theta + K(\widetilde X_t^{\eps} - \phi )\big)\big|^2,
  \end{aligned}
  \]
  which may be reorganized as follows (using~\eqref{eq:instanton} for estimating the time
  derivatives of~$\phi$ and~$\theta$):
  \[
  \begin{aligned}
    \alpha(t,\widetilde X_t^{\eps}) =&\ - |\sigma \theta|^2+ \frac{\eps}{2}
    D:K + \frac{1}{2} |\sigma \theta|^2
    + \frac{1}{2}\theta\cdot D  K (\widetilde X_t^{\eps} - \phi)
    + \frac{1}{2}\theta\cdot D  K^\mathrm{T} (\widetilde X_t^{\eps} - \phi)
    \\ & +\frac{1}{2} \big| \sigma K(\widetilde X_t^{\eps} - \phi)\big|^2 
   + \theta\big( b(\widetilde X_t^{\eps}) - b(\phi)\big)
  \\ & - \theta\cdot \nabla b(\phi) (\widetilde X_t^{\eps} - \phi)
   + \frac{1}{2}(\widetilde X_t^{\eps} - \phi)\cdot \dot{K} (\widetilde X_t^{\eps} - \phi)
  +\frac{1}{2}\big( b(\widetilde X_t^{\eps}) - b(\phi)\big)\cdot K (\widetilde X_t^{\eps} - \phi) 
  \\ & +\frac{1}{2}(\widetilde X_t^{\eps} - \phi)\cdot K^\mathrm{T} \big( b(\widetilde X_t^{\eps}) - b(\phi)\big)
   - \frac{1}{2}\theta \cdot D  K (\widetilde X_t^{\eps} - \phi)
  - \frac{1}{2}\theta \cdot D  K^\mathrm{T} (\widetilde X_t^{\eps} - \phi)
  \\ = & \ - \frac{1}{2} |\sigma \theta|^2 + \frac{\eps}{2}
    D:K
    +\frac{1}{2} \big| \sigma K(\widetilde X_t^{\eps} - \phi)\big|^2 
  + \theta\big( b(\widetilde X_t^{\eps}) - b(\phi)\big)
  - \theta\cdot \nabla b(\phi) (\widetilde X_t^{\eps} - \phi)
  \\ & + \frac{1}{2}(\widetilde X_t^{\eps} - \phi)\cdot \dot{K} (\widetilde X_t^{\eps} - \phi)
  +\frac{1}{2}\big( b(\widetilde X_t^{\eps}) - b(\phi)\big)\cdot K (\widetilde X_t^{\eps} - \phi) 
  +\frac{1}{2}(\widetilde X_t^{\eps} - \phi)\cdot K^\mathrm{T} \big( b(\widetilde X_t^{\eps}) - b(\phi)\big).
  \end{aligned}
  \]
  Inserting the expansion~\eqref{eq:Xtildehigh} then leads to
  \[
  \begin{aligned}
    \alpha(t,\widetilde X_t^{\eps}) =&\ - \frac{1}{2} |\sigma \theta|^2
    + \frac{\eps}{2} D:K
    +\frac{\eps}{2} |\sigma \zeta_t K|^2 + \sqrt{\eps}\theta\cdot
    \nabla b(\phi)\zeta_t
    + \eps \theta\cdot \nabla b(\phi) \beta_t + \frac{\eps}{2}\zeta_t\cdot \theta
    \nabla^2b(\phi)\zeta_t
    \\ & - \theta\cdot \nabla b(\phi)(\sqrt{\eps}\zeta_t + \eps \beta_t )
    + \frac{\eps}{2}\zeta_t \cdot \dot{K} \zeta_t
    +\frac{1}{2} \eps \zeta_t \cdot  \nabla b(\phi) K \zeta_t
    +\frac{1}{2} \eps \zeta_t \cdot K^\mathrm{T} \nabla b(\phi)\zeta_t.
  \end{aligned}
  \]
  Now, we may identify the terms of various orders in~$\sqrt{\eps}$ in the above
  equation. At leading order it remains~$-|\sigma\theta|^2/2$, which was expected. At
  order~$\sqrt{\eps}$ remains
  \[
   \theta\cdot \nabla b(\phi)\zeta_t - \theta\cdot \nabla b(\phi)\zeta_t
   = 0.
  \]
  We now turn to the terms of order~$\eps$ (excluding the term in~$D:K_t$ for now) that
  are given by
  \[
   \frac{1}{2} \zeta_t\cdot K D K \zeta_t
  + \theta\cdot \nabla b(\phi)\beta_t
  - \theta\cdot \nabla b(\phi)\beta_t + \frac{1}{2} \zeta_t\cdot \dot{K}\zeta_t
  + \frac{1}{2}\zeta_t \cdot \theta \nabla^2b(\phi)\zeta_t
  + \frac{1}{2}\zeta_t\cdot K^\mathrm{T} \nabla b(\phi)  \zeta_t
  + \frac{1}{2}\zeta_t\cdot  \nabla b(\phi)^\mathrm{T} K  \zeta_t.
  \]
  We see that the terms proportional to~$\beta_t$ cancel, while the quadratic
  product in~$\zeta_t$ factors out, so it remains:
  \begin{equation}
    \label{eq:Kproof}
  \dot{K}_t + K_t D K_t + K_t^\mathrm{T} \nabla b(\phi) + \nabla b^\mathrm{T}(\phi) K +  \theta \cdot\nabla^2 b(\phi),
  \end{equation}
  which is equal to~$0$ since~$(K_t)_{t\in[0,T]}$ is the solution to~\eqref{eq:EDOK}. Gathering
  the above results shows that~\eqref{eq:alphahigh} becomes, in the small~$\eps$ limit,
  \begin{equation}
    \label{eq:estimatealpha}
  \alpha(t,\widetilde X_t^{\eps}) = - \frac{1}{2} |\sigma \theta_t|^2 +
  \frac{\eps}{2} D: K_t+
  \mathrm{O} \big(\eps^{3/2}\big).
  \end{equation}

  In order to use the above estimates in~\eqref{eq:psiepsg}, we need~\eqref{eq:Xtildehigh}
  to hold. Such a perturbative formula holds for the dynamics~\eqref{eq:Xtilde} provided it
  starts from the correct initial condition at time~$t$ when computing~\eqref{eq:psiepsg} --
  in other words, if~$x$ is far from~$\phi_t$, the error may well be large. One way to solve this
  problem is to note that~\eqref{eq:Xtildehigh} is actually satisfied at any
  time when the process is started at
  time~$t$ from the value~$\phi_t$ of the instanton at that time.
  Introducing the shorthand notation
  \[
Y_T = \e^{ \frac{1}{\eps}( f(\widetilde X_T^{\eps}) - g^1(T,\widetilde X_T^{\eps}) )
      +\frac{1}{\eps}\int_t^T \alpha(s,\widetilde X_s^\eps)\,ds},
\]
we may thus rewrite~\eqref{eq:psiepsg} as
  \begin{equation}
    \label{eq:psiepsgfrac}
      \psi_{\eps}(t,x) 
   = \e^{\frac{g^1(t,x)}{\eps} } \E_{t,\phi_t}\left[
    \e^{ \frac{1}{\eps}( f(\widetilde X_T^{\eps}) - g^1(T,\widetilde X_T^{\eps}) )
      +\frac{1}{\eps}\int_t^T \alpha(s,\widetilde X_s^\eps)\,ds}
    \right]\frac{\E_{t,x}[Y_T]}{\E_{t,\phi_t}[Y_T]}.
  \end{equation}
  In the first expectation starting from~$(t,\phi_t)$ we can then perform the expansions as above.

  Plugging the estimates~\eqref{eq:hg} and~\eqref{eq:estimatealpha} into the first
  expectation in~\eqref{eq:psiepsgfrac}, we obtain
  \[
    \psi_{\eps}(t,x) = \e^{\frac{g^1(t,x)}{\eps} } \E_{t,\phi_t}\left[
      \e^{ \frac{1}{\eps}\big( f( \phi_T) -\frac{1}{2}\int_t^T
        |\sigma\theta_s|^2\,ds
        + \frac{\eps}{2}\int_t^T D: K_s\,ds 
        +   \mathrm{O} (\eps^{3/2} ) \big)}
      \right]\frac{\E_{t,x}[Y_T]}{\E_{t,\phi_t}[Y_T]}.
    \]
    Taking the logarithm and multiplying by~$\eps$ then leads to
    \[
    \begin{aligned}
    g_{\eps}(t,x)  =&\ \eps \log\psi_\eps(t,x)
    =& g^1(t,x) &+ f(\phi_T) -\frac{1}{2}\int_t^T
        |\sigma\theta_s|^2\,ds + \frac{\eps}{2}\int_t^T D: K_s\,ds 
        \\ & & & + \eps \log \E_{t,\phi_t}
        \left[
          \e^{ \frac{1}{\eps} \mathrm{O} (\eps^{3/2} ) }\right]
        +\eps\left[ \log \E_{t,x}[Y_T] - \log \E_{t,\phi_t}[Y_T] 
          \right].
        \end{aligned}
        \]
        Assuming that $(t,x)\to\log \E_{t,x}[Y_T]$ is smooth, we consider the
        Taylor expansion in~$x$ around~$\phi_t$ for the difference of logarithms in addition to the
        small~$\eps$ limit.
    As a result, the optimal control~\eqref{eq:gopt} admits the following expansion
    \[
    g_{\eps}(t,x) = g^1(t,x) + f(\phi_T) -\frac{1}{2}\int_t^T
    |\sigma\theta_s|^2\,ds + \frac{\eps}{2}\int_t^T D: K_s\,ds 
        +   \mathrm{o} (\eps) + \eps\mathrm{o}\big(x - \phi_t\big),
    \]
    in the small~$\eps$ regime and for~$x$ close to~$\phi_t$,
    where~$g^1$ is defined in~\eqref{eq:g2}. This provides the desired result.

\bibliographystyle{abbrv}
\bibliography{bib}

\end{document}